\newif\ifarxiv
\arxivfalse       

\ifarxiv
\documentclass[
  preprint,
  amsmath,amssymb,
  aps,
  prl,
]{revtex4-2}
\else
\documentclass[
  reprint,
  amsmath,amssymb,
  aps,
  prl,
  longbibliography
]{revtex4-2}
\fi

\usepackage{graphicx}   
\usepackage{dcolumn}    
\usepackage{bm}         
\usepackage{xcolor}
\usepackage{physics}
\usepackage{siunitx}
\usepackage[normalem]{ulem} 
\usepackage{booktabs}
\usepackage{hyperref}   
\begin{document}


\title{Coherent Phonon Generation from Prethermal Carrier Populations in a Semiconductor}


\author{Mingyu Shi}
\email{minshi@phys.ethz.ch}
\affiliation{Institute for Quantum Electronics, ETH Zurich, Switzerland}

\author{Matteo Savoini}
\affiliation{Institute for Quantum Electronics, ETH Zurich, Switzerland}

\author{Patrick Friebel}
\affiliation{
Max-Planck-Institut f\"ur Kernphysik, Heidelberg, Germany}

\author{Andrea Annunziata}
\affiliation{
Max-Planck-Institut f\"ur Kernphysik, Heidelberg, Germany}

\author{Thomas Metzger}
\affiliation{Institute for Quantum Electronics, ETH Zurich, Switzerland}

\author{Yunpei Deng}
\affiliation{Center for Photon Science, Paul Scherrer Institute, Switzerland}

\author{Laura Cattaneo}
\affiliation{
Max-Planck-Institut f\"ur Kernphysik, Heidelberg, Germany}

\author{Steven L. Johnson}
\email{johnson@phys.ethz.ch}
\affiliation{Institute for Quantum Electronics, ETH Zurich, Switzerland}
\affiliation{Center for Photon Science, Paul Scherrer Institute, Switzerland}

\date{\today}

\begin{abstract}
Ultrafast laser excitation of carriers in semiconductors often results in the creation of fully symmetric coherent optical phonons, typically understood in terms of a thermalized transient carrier population. We test this understanding with systematic measurements and first-principles calculations of the photon-energy dependence of ultrafast-laser-driven coherent phonon generation in tellurium, using excitation wavelengths around the electronic band gap. 
We observe that, contrary to the expectations of models assuming a thermalized carrier population, the coherent amplitude of the fully symmetric $A_1$ mode changes only weakly when tuning wavelength across the band gap. The measured spectral dependence is instead reproduced by a transition-resolved prethermal force weighted by optical matrix elements and band- and momentum-dependent $A_1$ deformation potentials.  Fitting this model to the data yields an excited transition state lifetime of a few femtoseconds. These results identify transition selectivity, rather than a thermalized total carrier density, as a key factor controlling coherent-phonon generation in semiconductors excited near a band gap.
\end{abstract}

\maketitle

Coherent phonons created by ultrafast laser interactions have attracted increasing attention as a means of controlling material properties on picosecond timescales, enabling the manipulation and even engineering of collective excitations through nonlinear and parametric coupling \cite{forst2011nonlinear,cartella2018parametric,michael2022generalized,haque2024terahertz}. There are a variety of physical mechanisms for generating coherent phonons, each with important limitations. One important class of mechanisms acts via electron--phonon coupling and is often relevant to driving large-amplitude fully symmetric coherent phonon modes in semiconductors using light frequencies near and above the electronic band gap.

There are several related but conceptually distinct descriptions of this mechanism in the literature. One is impulsive stimulated Raman scattering (ISRS), which assumes nonresonant driving of electronic transitions and is typically applicable at wavelengths for which the material is transparent~\cite{Nelson1985ISRS,YamadaPRB2019}. For ISRS, coherent phonon excitation can be described in a semiclassical Raman picture, in which the laser interaction results in a transient effective force on the phonon mode that follows the intensity profile of the laser pulse~\cite{merlin1997generating}. Another framework for describing coherent phonon generation is the so-called displacive excitation of coherent phonons (DECP) theory, in which photoexcited carriers are assumed to rapidly thermalize to a Fermi--Dirac distribution with an elevated effective temperature~\cite{cheng1991mechanism,zeiger1992theory}. This results in a new effective lattice potential with a quasi-equilibrium structural configuration characterized by displacements along fully symmetric phonon-mode coordinates. Assuming that this displacement of the quasi-equilibrium structure occurs on a timescale much shorter than the periods of the phonon modes, the result is an effective step-like force on the fully symmetric modes that depends on the transient carrier density and temperature. DECP theory is typically applied to semiconductors when the laser excitation is above the band gap and can be modeled using density functional theory (DFT)~\cite{LakehalPRB2019}.

As a step toward unifying these descriptions of coherent phonon generation, Stevens \textit{et al.} introduced a theory of impulsive Raman scattering for both resonantly and nonresonantly driven electronic transitions, using a generation tensor whose imaginary part contains a contribution from resonant electronic excitations \cite{stevens2000ultrafast,stevens2002coherent}. Like DECP theory, it predicts an enhanced, step-like coherent-phonon driving force from laser-driven excitations across an electronic band gap. Unlike DECP theory, it considers the electronic transitions driven by the laser and does not account for carrier thermalization.

Despite the ubiquity of coherent phonon phenomena in ultrafast spectroscopy, there have been limited efforts to test experimentally the predictions of these models for near-gap excitation.
Previous studies have focused mainly on semimetals such as antimony and bismuth \cite{garrett1996coherent,melnikov2011generation}, where the absence of a well-defined band gap and the rapid redistribution of the excited electronic population complicate models \cite{johnson2008nanoscale,timrov2012thermalization,melnikov2013ultrafast,sheu2013free}. A gapped semiconductor provides a cleaner setting because photoexcited carriers typically persist with a well-defined density for many phonon periods \cite{zhuo2024ultrafast}.

Here, we combine photon-energy- and fluence-dependent pump-probe reflectivity measurements across the fundamental band gap of semiconducting tellurium (Te) with first-principles calculations to test these different models of coherent phonon generation. We observe that the amplitude of the coherently excited $A_1$ mode increases only weakly as the excitation photon energy increases across the band gap, contrary to the predictions of existing theories. Our first-principles calculations instead attribute the measured spectral dependence to the effective force exerted by the initially excited, prethermal carrier population. This force is determined by the optical matrix elements and the band- and momentum-dependent $A_1$ deformation potentials.

We performed the measurements at room temperature on a bulk $c$-cut Te single crystal grown by the Czochralski method with a measured optical band gap of approximately $0.325~\mathrm{eV}$~\cite{SM}. The time-resolved (pump--probe) reflectivity measurements were collected using Ti:sapphire-based regeneratively amplified laser systems operating at $1~\mathrm{kHz}$ (for full details see ref.~\onlinecite{SM}). An optical parametric amplifier (OPA) creates tunable pump pulses close to Te bandgap, while a horizontally polarized $800~\mathrm{nm}$ pulse serves as the probe. We employed two different systems with optimized outputs in different spectral ranges. The sole difference between them is the fundamental pulse duration (100~fs and 35~fs FWHM). The $100$-fs system generates vertically polarized mid-IR pulses at $0.275$, $0.310$, $0.325$, $0.339$, and $0.355~\mathrm{eV}$.  
We performed experiments at higher pump photon energies on the system with pulse duration of $35~\mathrm{fs}$ FWHM. For this setup, the pump is tuned to $0.358$, $0.429$, $0.470$, and $0.485~\mathrm{eV}$. The experimental geometry is similar in both setups, with the probe incident nearly normal to the sample and focused to a few tens of $\mu$m FWHM spot. The pump pulses are focused to $>250$~$\mu$m FWHM, ensuring nearly homogeneous excitation over the probed area. $F_{\mathrm{inc}}$ denotes the peak incident fluence calculated from the pulse energy and measured Gaussian-beam spot sizes.

\begin{figure}[t]
    \centering
    \includegraphics[width=\linewidth]{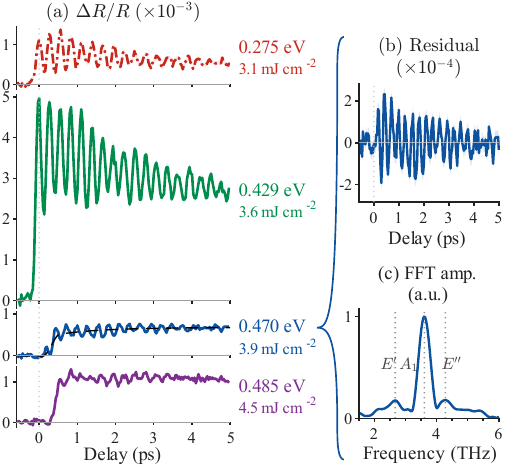}
    \caption{Representative transient reflectivities and coherent-phonon extraction. (a) Unscaled $\Delta R/R$ traces on a common amplitude scale; dash-dotted curves denote the longer-pulse setup. For photon energies $\leq0.429~\mathrm{eV}$, $t=0$ marks the first transient maximum; higher-energy traces retain the $0.429$-eV delay zero. The black dashed curve is the $0.470$-eV background fit~\cite{SM}. (b) Background-subtracted residual for the $0.470$~eV pump data; shading denotes the standard error of the mean (SE). (c) Fourier spectrum for $0\leq t\leq5~\mathrm{ps}$ for the data in (b), with the $E^\prime$, $A_1$, and $E^{\prime\prime}$ modes marked.}
    \label{pic:exp_results}
\end{figure}

Fig.~\ref{pic:exp_results}(a) shows representative transient-reflectivity traces from both setups. Up to $0.429~\mathrm{eV}$, the pump induces a prompt rise in $\Delta R/R$ followed by damped oscillations superimposed on a slowly decaying background, characteristic of photoexcited-carrier relaxation in semiconductors~\cite{knox1986femtosecond,sabbah2002femtosecond}. At higher photon energies,  both the oscillatory and non-oscillatory responses change qualitatively. At $0.470~\mathrm{eV}$, the overall signal is approximately five times smaller than at $0.429~\mathrm{eV}$ despite the similar pump fluence, and the background initially grows on a time scale similar to that of an oscillation period and then remains nearly constant over a $5~\mathrm{ps}$ window. At $0.485~\mathrm{eV}$, the long-lived offset from the background is similar in shape but larger, whereas the oscillations are somewhat smaller and noisy. 

To isolate the coherent oscillations, we subtract a photon-energy-dependent background~\cite{SM}, illustrated by the black dashed fit to the $0.470$-eV trace in Fig.~\ref{pic:exp_results}(a). The resulting residual is shown in Fig.~\ref{pic:exp_results}(b). Its Fourier spectrum over $0\leq t\leq5~\mathrm{ps}$ [Fig.~\ref{pic:exp_results}(c)] is dominated by the $A_1$ mode at $3.6~\mathrm{THz}$, with weaker $E^\prime$ and $E^{\prime\prime}$ peaks at approximately $2.76$ and $4.2~\mathrm{THz}$, respectively~\cite{torrie1970raman,pine1971raman}.

\begin{figure}[tbp]
    \centering
    \includegraphics[width=\linewidth]{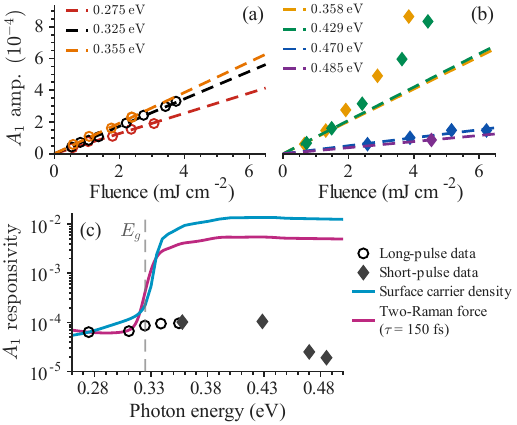}
   \caption{Fluence dependence and spectral evolution of the coherent $A_1$ response. (a),(b) Peak Fourier amplitudes versus incident fluence for long-pulse (open circles) and short-pulse (filled diamonds) data; dashed lines show linear fits. (c) Responsivity versus photon energy compared with surface photocarrier density at fixed fluence (cyan) and two-Raman force magnitude at $\Omega_{A_1}$ (magenta), calculated from $a_{TT}$ for a $150$-fs pulse. Both curves are scaled to the lowest-energy responsivity. The gray dashed line marks $E_g\simeq0.325~\mathrm{eV}$.}
    \label{pic:a1_fluence_and_two_raman}
\end{figure}

Figs.~\ref{pic:a1_fluence_and_two_raman}(a) and \ref{pic:a1_fluence_and_two_raman}(b) show the peak spectral amplitude of the $A_1$ mode as a function of fluence for the different pump photon energies. Below the band gap (Fig.~\ref{pic:a1_fluence_and_two_raman}(a)), the dependence of the A\(_1\) amplitude on pump fluence is well-described by a linear proportionality, consistent with a non-resonant Raman process or displacive excitation from weak absorption~\cite{SM}, and does not suggest contributions from  multiphoton processes. Fig.~\ref{pic:a1_fluence_and_two_raman}(b) shows that the $0.358$- and $0.429$-eV responses are approximately linear at low fluences but deviate from linearity above approximately $2~\mathrm{mJ\,cm^{-2}}$, whereas the $0.470$-eV response remains mostly linear over the measured fluence range. At $0.485~\mathrm{eV}$, the limited signal-to-noise ratio and available pump power allowed us to measure at only one fluence, so the line through the origin defines an approximate effective slope rather than establishing linearity. We define the responsivity $a$ of the $A_1$ mode to the pump excitation as the slope of the linear fits shown in Figs.~\ref{pic:a1_fluence_and_two_raman}(a,b). Fig.~\ref{pic:a1_fluence_and_two_raman}(c) shows that the responsivity is weakly enhanced upon crossing the band gap and then decreases strongly at higher photon energies.

To understand our results, we first turn to the predictions of DECP theory. In this framework, the excited carrier density and  temperature determine the amplitude of the coherent phonon~\cite{zeiger1992theory}. In the linear absorption regime, we estimate the surface photocarrier density as
\begin{equation}
N_{\mathrm{surf}}=\frac{F_{\mathrm{inc}}(1-R)\alpha(\mathcal{E})}{\mathcal{E}},\label{eq:Nsurf}
\end{equation}
where $R$ is the pump reflectivity and $\alpha(\mathcal{E})$ the absorption coefficient~\cite{SM}.  For all pump photon energies measured, $R \approx 0.44$.  The penetration depth \(1/\alpha\) varies dramatically from $1.46\,\mu\textrm{m}$ to $513\,\mu\textrm{m}$, but remains larger than the probe attenuation length of $61\,\textrm{nm}$~\cite{SM}. The excitation density is therefore well-approximated by eq.~\ref{eq:Nsurf}. 
The dependence of \(N_\textrm{surf}\) on the pump photon energy is shown in Fig.~\ref{pic:a1_fluence_and_two_raman}(c), where we compare it to the spectral dependence of the measured responsivity.  The excited carrier density shows a rapid increase (100$\times$) across the bandgap, in contrast to the measured responsivity.  

This comparison suggests that DECP theory is not an accurate model for our data. To test this further we perform DFT calculations using \textsc{Quantum ESPRESSO}~\cite{giannozzi2009quantum,giannozzi2017advanced,giannozzi2020quantum,SM}. We model photoexcitation within a constrained two-chemical-potential framework~\cite{marini2021lattice} by transferring a prescribed fraction of electrons from the valence to the conduction manifold. The electron and hole occupations are described by separate Fermi-Dirac distributions at a common carrier temperature of $k_{\mathrm{B}}T=0.05~\mathrm{eV}$, which provides a simplified quasi-equilibrium description of a thermalized electron--hole plasma. The internal coordinates are then relaxed at fixed ground-state lattice parameters. We define $u_{A_1}$ as the unweighted Cartesian amplitude along the normalized three-atom $A_1$ eigendisplacement vector, with $u_{A_1}>0$ corresponding to expansion of the Te helix. The resulting displacement agrees closely with previous first-principles calculations \cite{tangney1999calculations,tangney2002density, SM}. For the $0.485$-eV measurement at $F_{\mathrm{inc}}=4.54~\mathrm{mJ\,cm^{-2}}$,  $N_{\mathrm{surf}}$ corresponds to an excitation fraction of approximately $0.14\%$ of the valence-electron population, which in our DFT model yields a quasi-equilibrium displacement of $u_{A_1}\approx0.8~\mathrm{pm}$. Varying the carrier temperature from $k_{\mathrm{B}}T=0.5$ to $0.02~\mathrm{eV}$ changes $u_{A_1}$ by at most  $0.1~\mathrm{pm}$, indicating only a weak temperature dependence. The calculated displacement is comparable to the $1.15~\mathrm{pm}$ shift measured by ultrafast x-ray diffraction at an excitation fraction of $0.2\%$~\cite{johnson2009full} achieved using a $1.5$ eV photon energy pump, supporting the presence of a moderate carrier-induced structural shift. 

Under DECP theory, the quasi-equilibrium displacement of the $A_1$ mode is nearly identical to the amplitude of the coherent phonon. Our transient reflectivity measurements measure the coherent phonon amplitude indirectly via the magnitude of the reflectivity oscillations. With a pump photon energy and fluence nominally identical to that used in the x-ray diffraction study of ref.~\onlinecite{johnson2009full}, Hunsche et al. report $800$-nm reflectivity probe measurements and observe $A_{1}$ oscillations of $\Delta R/R\approx2\%$~\cite{hunsche1995impulsive}. This is $50$ times larger than the amplitude of $0.04\%$ measured here at the same probe wavelength at similar excited carrier densities [Fig.~\ref{pic:exp_results}(a), bottom curve]. Although this comparison is approximate because of differences in the pump and probe penetration depths, it indicates that DECP theory severely overestimates the observed coherent phonon amplitude.

As an alternative to the thermalized-carrier DECP model, we next turn to the two-Raman-tensor formalism of Stevens \textit{et al.}~\cite{stevens2002coherent}.  Following this work, we apply the assumptions of momentum-independent electron--phonon coupling and infinite electronic-state lifetimes. After averaging the energy-dependent Raman driving component over the finite pump bandwidth, Eqs.~(9) and (10) of ref.~\onlinecite{stevens2002coherent} give the temporal Fourier transform of the effective force, up to an overall energy-independent constant, as
\begin{equation}
\begin{aligned}
F_{\mathrm{2R}}(\Omega;\mathcal{E}_0,\tau)
&\propto (1-R)\widetilde{I}_{\mathrm{inc},\tau}(\Omega;\mathcal{E}_0)
\Bigg[
\left\langle\frac{d\epsilon_1}{d\mathcal{E}}\right\rangle_{\mathcal{E}_0,\tau} \\
&\quad+\mathrm{i}\Bigg(
\left\langle\frac{d\epsilon_2}{d\mathcal{E}}\right\rangle_{\mathcal{E}_0,\tau}
+\frac{2}{\hbar\Omega}
\left\langle\epsilon_2\right\rangle_{\mathcal{E}_0,\tau}
\Bigg)\Bigg].
\end{aligned}
\label{eq:Stevens}
\end{equation}
where, for any energy-dependent function $g$,
\begin{equation}
\left\langle g\right\rangle_{\mathcal{E}_0,\tau}\equiv\int d\mathcal{E}\,w_{\tau}(\mathcal{E};\mathcal{E}_0)g(\mathcal{E}),\;\;\:
\int d\mathcal{E}\,w_{\tau}(\mathcal{E};\mathcal{E}_0)=1.
\label{eq:bandwidth_average}
\end{equation}
Here, $\Omega$ is the angular frequency, $\mathcal{E}_0$ is the central pump photon energy, $\epsilon=\epsilon_1+\mathrm{i}\epsilon_2$ is the complex in-plane dielectric function, and $w_{\tau}(\mathcal{E};\mathcal{E}_0)$ is the normalized spectral-intensity distribution of a transform-limited Gaussian pump pulse with intensity FWHM $\tau$. The factor $\widetilde{I}_{\mathrm{inc},\tau}(\Omega;\mathcal{E}_0)$ is the Fourier component of the incident pump-intensity envelope. Eq.~\ref{eq:Stevens} generally yields a complex spectral force: its real part corresponds to a time-domain force that follows the pump-intensity envelope, whereas the imaginary part gives a force term proportional to the time-integral of the intensity envelope (i.e. displacive). We estimate $\epsilon(\mathcal{E})$ from literature data and spectroscopic measurements on our sample~\cite{SM}, retaining the $d\epsilon_2/d\mathcal{E}$ term because $\epsilon_2$ varies rapidly near the Te band gap. 
We then evaluate the two-tensor responsivity \(a_{TT} \propto |F(\Omega_0;\mathcal{E}_0,\tau)|/F_\textrm{inc}\) for each pump photon energy \(\mathcal{E}_0\) with $\tau=150~\mathrm{fs}$, where we set \(\Omega_0/2\pi  = 3.6\)~THz. The result is shown as a solid curve in  Fig.~\ref{pic:a1_fluence_and_two_raman}(c) that is normalized to match the experimentally measured responsivity at the lowest measured pump photon energy.  Similar to the surface carrier density, the imaginary absorptive contribution $2\epsilon_2/(\hbar\Omega)$ causes a sharp rise $(10\times)$ as the photon energy crosses the band gap, followed by a flat response at higher photon energies. This clearly disagrees with the spectral dependence of the measured responsivity.

\begin{figure}[tbp]
    \centering
    \includegraphics[width=\linewidth]{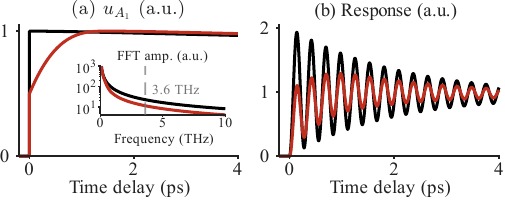}
    \caption{Effects of finite-time carrier thermalization on coherent phonon dynamics. (a) Instantaneous (black) and partially prompt, slowly rising (red) equilibrium shifts with equal maxima. A weak long-time decay ensures a well-defined Fourier transform. Inset: Driving-force spectra over the full time window; the dashed line marks the 3.6-THz $A_1$ frequency. (b) Corresponding phonon trajectories with identical frequency and damping.}
    \label{pic:decp_force_toy_model}
\end{figure}

To explain the deviation of the responsivity from these models, we propose an alternative scenario where the effective force on the phonon coordinate is strongly influenced by the thermalization kinetics of the excited carriers. The pump pulse initially creates a highly nonthermal carrier distribution which takes some time to thermalize. If the initially populated excited state couples differently to the A$_1$ displacement than does the final thermalized state, the effective force on the A$_1$ mode will evolve in time after the initial excitation. If the thermalization time is comparable to or longer than the period of the A\(_1\) mode, the result will be a suppression of the coherent phonon response relative to the displacement after carrier thermalization. This idea is illustrated schematically in Fig.~\ref{pic:decp_force_toy_model}, which shows a comparison of the effective force and mode response of the thermalized-DECP theory to a model that takes into account a relatively weaker effective force from pre-thermalized carriers that slowly develops as thermalization occurs. Our data for pump energies above the bandgap provide some qualitative support for this idea:  for example, at  $0.470~\mathrm{eV}$ (see Fig.~\ref{pic:exp_results}(a)) we can see that the non-oscillatory background is not step-like, but increases gradually on a time scale of approximately 0.5 ps before reaching a time-independent plateau.  We observe similar behavior at $0.485$~eV, and to a lesser extent there is a slow component to the background even for lower pump photon energies.

To estimate more precisely the effective force caused by carrier population changes immediately after excitation and before significant carrier-carrier scattering or ionic motion, we develop a model based on first-principles DFT. For instantaneous electronic occupations $\{f_{n\mathbf{k}}(t)\}$ with band index $n$ and momentum $\mathbf{k}$, the effective force along the $A_1$ coordinate is
\begin{equation}
\mathcal{F}_{A_1}\!\left[\{f_{n\mathbf{k}}\}\right]=-\left.\frac{\partial E_{\mathrm{BO}}\!\left[\{f_{n\mathbf{k}}\},Q_{A_1}\right]}{\partial Q_{A_1}}\right|_{Q_{A_1}=0},
\label{eq:force_general}
\end{equation}
where $E_{\mathrm{BO}}$ is the Born-Oppenheimer energy for one unit cell. We assume that the occupations are independent of $Q_{A_1}$ to leading order. By definition, this force vanishes in the relaxed ground state. Let $E_{n\mathbf{k}}$ be the energy of an electron in band $n$ at momentum $\mathbf{k}$. 
Letting $f^{(0)}_{n\mathbf{k}}$ denote the ground-state occupation, we expand the force to first order in the occupation change $\delta f_{n\mathbf{k}}=f_{n\mathbf{k}}-f^{(0)}_{n\mathbf{k}}$ and obtain
\begin{equation}
\mathcal{F}_{A_1} \approx -\sum_{n,\mathbf{k}}\,\delta f_{n\mathbf{k}}(t)\,g_n(\mathbf{k}) ,
\label{eq:force_linearized}
\end{equation}
where $g_n(\mathbf{k})=\left.\partial E_{n\mathbf{k}}/\partial Q_{A_1}\right|_{Q_{A_1}=0}$ is the band- and momentum-resolved electron--phonon coupling to the $A_1$ coordinate. Let $\delta f_{cv}(\mathbf{k})$ be the change in occupations caused by a vertical transition at momentum $\mathbf{k}$ from an initially fully occupied valence band $v$ to an initially unoccupied conduction band $c$. Assuming that all changes in electronic occupation are caused by such transitions, Eq.~\eqref{eq:force_linearized} can be written as
\begin{equation}
\mathcal{F}_{A_1}=-\sum_{\mathbf{k},v,c}\,\delta f_{cv}(\mathbf{k})\left[g_c(\mathbf{k})-g_v(\mathbf{k})\right],
\label{eq:force_transition}
\end{equation}
where $v$ and $c$ are summed over all valence and conduction bands, respectively. We approximate the transition-resolved population generated by a pump of photon energy $\mathcal{E}$ in the weak-excitation regime as
\begin{equation}
\delta f_{cv}(\mathbf{k};\mathcal{E})\propto F_{\mathrm{inc}}\left|\mathbf{e}\cdot\mathbf{r}_{cv}(\mathbf{k})\right|^2G\!\left[\mathcal{E}-\Delta E_{cv}(\mathbf{k}),\sigma\right],
\label{eq:population_ansatz}
\end{equation}
where $F_{\mathrm{inc}}$ is the incident pump fluence, $\mathbf{e}$ is the pump-polarization vector, $\mathbf{r}_{cv}(\mathbf{k})$ is the interband dipole matrix element, $\Delta E_{cv}(\mathbf{k})=E_{c\mathbf{k}}-E_{v\mathbf{k}}$, and $G$ is a normalized Gaussian with root-mean-square width $\sigma$. Combining Eqs.~\eqref{eq:force_transition} and \eqref{eq:population_ansatz}, and defining the polarization-resolved optical matrix element $M_{cv}(\mathbf{k})=\mathbf{e}\cdot\mathbf{r}_{cv}(\mathbf{k})$ and the transition-resolved coupling difference $\Delta g_{cv}(\mathbf{k})=g_c(\mathbf{k})-g_v(\mathbf{k})$, yields the prethermal force spectrum
\begin{equation}
\begin{aligned}
\mathcal{F}_{A_1}(\mathcal{E})
&\propto -F_{\mathrm{inc}}
\sum_{\mathbf{k},v,c} 
\Delta g_{cv}(\mathbf{k})\\
&\quad \times \left|M_{cv}(\mathbf{k})\right|^2 
G\!\left[\mathcal{E}-\Delta E_{cv}(\mathbf{k}),\sigma\right].
\end{aligned}
\label{eq:impulsive_force}
\end{equation}
For comparison, the corresponding dipole-weighted joint density of states (JDOS) is
\begin{equation}
\mathcal{J}(\mathcal{E})
\propto
\sum_{\mathbf{k},v,c} 
\left|M_{cv}(\mathbf{k})\right|^2
G\!\left[\mathcal{E}-\Delta E_{cv}(\mathbf{k}),\sigma\right].
\label{eq:dipole_weighted_jdos}
\end{equation}
Each transition is therefore weighted by both its transition probability and its coupling to the $A_1$ mode (magnitude and sign). Consequently, different transition channels may reinforce or cancel one another in the total force even when the overall absorption remains strong.

\begin{figure}[tbp]
    \centering
    \includegraphics[width=\linewidth]{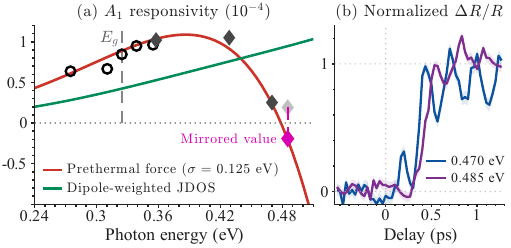}
    \caption{(a) Measured $A_1$ responsivities (symbols), scaled prethermal force (red), and dipole-weighted JDOS (green) for $\sigma=0.125~\mathrm{eV}$. Circles and diamonds denote long- and short-pulse data. Light-gray and magenta diamonds show the measured and sign-inverted $0.485$-eV values, joined by a magenta dashed line. The vertical gray dashed line marks $E_g\simeq0.325~\mathrm{eV}$; positive force contracts the Te helix. (b) $0.470$-eV (blue) and $0.485$-eV (purple) traces from Fig.~\ref{pic:exp_results}, normalized to their $4$--$5~\mathrm{ps}$ plateaus; shaded bands show SE.}
    \label{pic:a1_responsivity_prethermal_force}
\end{figure}

Fig.~\ref{pic:a1_responsivity_prethermal_force}(a) shows the responsivity derived from the prethermal force calculated at photon energies from $0.24~\mathrm{eV}$ to $0.51~\mathrm{eV}$ using a DFT model of the electronic band structure~\cite{SM}, together with the dipole-weighted JDOS from Eq.~\ref{eq:dipole_weighted_jdos} and the measured $A_1$ responsivity. Here, the broadening parameter $\sigma$ and the overall scaling factor of the prethermal force are fit to the data. The dipole-weighted JDOS in Fig.~\ref{pic:a1_responsivity_prethermal_force}(a) is evaluated with the same $\sigma$ and rescaled separately for display.
We find that $\sigma=0.125~\mathrm{eV}$ best reproduces the spectral shape of the responsivity, assuming that the prethermal force alone determines the coherent phonon amplitude. By contrast, the dipole-weighted JDOS increases monotonically throughout the investigated energy range, demonstrating that the suppression arises from the changing balance of transition-resolved deformation potentials rather than from a reduction of the available optical phase space. This explains why the coherent $A_1$ response can be nearly completely suppressed while the material remains strongly absorbing and supports a prompt, band- and momentum-selective driving mechanism rather than one governed solely by the total photocarrier density. We note that the calculated prethermal force changes sign near $0.47~\mathrm{eV}$. A change in the sign would manifest in the experiment as a $\pi$ phase inversion of the oscillatory response. Fig.~\ref{pic:a1_responsivity_prethermal_force}(b) compares the reflectivity dynamics at $0.470$ and $0.485~\mathrm{eV}$, just below and above the predicted zero crossing. Although the $0.485$-eV oscillations are close to the noise floor, their phase appears inverted relative to those at $0.470~\mathrm{eV}$. This interpretation remains tentative due to possible uncertainties in pump--probe timing shifts~\cite{SM}. Nevertheless, the prethermal-force spectrum in Fig.~\ref{pic:a1_responsivity_prethermal_force}(a) closely follows the sign-inverted $0.485$-eV data and accounts for its weaker responsivity relative to that at $0.470~\mathrm{eV}$.

The parameter $\sigma$ represents both the actual bandwidth of the pulses (approximately $0.01~\mathrm{eV}$~\cite{SM}) and the lifetime broadening of the excited transitions. The latter appears to be the dominant factor and  corresponds to an excited state lifetime of $\tau_{\mathrm{coh}}\sim\hbar/\sigma\approx5~\mathrm{fs}$. 
Strong carrier-carrier scattering resulting in similarly short excited state lifetimes has been reported in other highly excited semiconductors at comparable excitation densities~\cite{worle2021ultrafast,schultze2014attosecond}.

Our measurements and calculations show that even when excited carrier populations survive for many phonon periods, momentum- and band-dependent variations in the electron--phonon coupling strength can substantially modify the magnitude and character of the coherent phonon response.  Recent work using  high-energy XUV pulses to excite coherent phonons in Te suggests that the band-dependence of the phonon coupling can in some cases be even stronger for extremely deep hole populations~\cite{dogadov2026xray}.  
Strong momentum dependence of the electron--phonon coupling is not unique to Te, but has also been observed experimentally in lead-halide perovskites and $\mathrm{WSe_2}$~\cite{qu2025mode,waldecker2017momentum}. Despite this, the effects of prethermal populations have not yet been sufficiently accounted for in widely applied models of coherent phonon generation, leading to inaccurate estimates of coherent phonon amplitudes and phases. This work lays the foundation for a more general model to improve predictions and understanding of lattice dynamics.

\begin{acknowledgments}
This research was supported by the Swiss National Science Foundation through Grant 200021-228373.
\end{acknowledgments}
\bibliography{references_PRL_cleaned}
\bibliographystyle{apsrev4-2}
\end{document}


\title{Supplemental Material for ``Coherent Phonon Generation from Prethermal Carrier Populations in a Semiconductor''}

\author{Mingyu Shi}
\email{minshi@phys.ethz.ch}
\affiliation{Institute for Quantum Electronics, ETH Zurich, Switzerland}
\author{Matteo Savoini}
\affiliation{Institute for Quantum Electronics, ETH Zurich, Switzerland}
\author{Patrick Friebel}
\affiliation{Max-Planck-Institut f\"ur Kernphysik, Heidelberg, Germany}
\author{Andrea Annunziata}
\affiliation{Max-Planck-Institut f\"ur Kernphysik, Heidelberg, Germany}
\author{Thomas Metzger}
\affiliation{Institute for Quantum Electronics, ETH Zurich, Switzerland}
\author{Yunpei Deng}
\affiliation{Center for Photon Science, Paul Scherrer Institute, Switzerland}
\author{Laura Cattaneo}
\affiliation{Max-Planck-Institut f\"ur Kernphysik, Heidelberg, Germany}
\author{Steven L. Johnson}
\email{johnson@phys.ethz.ch}
\affiliation{Institute for Quantum Electronics, ETH Zurich, Switzerland}
\affiliation{Center for Photon Science, Paul Scherrer Institute, Switzerland}
\date{\today}
\maketitle

\section{Experimental methods}

\subsection{Laser systems and pump--probe setup}

We performed room-temperature pump--probe reflectivity measurements on a bulk $c$-cut Te single crystal grown by the Czochralski method using two Ti:sapphire-based regenerative-amplifier systems operating at $1~\mathrm{kHz}$. In both systems, difference-frequency generation (DFG) from an optical parametric amplifier (OPA) produced tunable mid-infrared pump pulses. A horizontally polarized $800$-nm probe measured the reflectivity at nearly normal incidence.

The first system used $100$-fs FWHM amplifier pulses to drive the OPA and generated vertically polarized pumps at $4511$, $3994$, $3820$, $3660$, and $3495~\mathrm{nm}$ ($0.275$, $0.310$, $0.325$, $0.339$, and $0.355~\mathrm{eV}$, respectively). The pump spot measured approximately $400\times400~\mu\mathrm{m}$ FWHM, enclosing a $62~\mu\mathrm{m}$ FWHM probe spot. The pump was modulated at $500~\mathrm{Hz}$, and a photodiode detected the transient reflectivity change $\Delta R/R$.

The second system used $35$-fs FWHM amplifier pulses and generated pumps at measured peak wavelengths of $3466$, $2893$, $2638$, and $2555~\mathrm{nm}$ ($0.358$, $0.429$, $0.470$, and $0.485~\mathrm{eV}$, respectively). Its approximately $330\times200~\mu\mathrm{m}$ FWHM pump spot enclosed a probe spot of order $10~\mu\mathrm{m}$ FWHM. These larger pump spots provided nearly homogeneous excitation over the probed area in both setups. 

In both measurement sets, the horizontal and vertical pump-spot diameters were determined by knife-edge scans at the sample position. For an elliptical Gaussian beam, the peak incident fluence is calculated as
\begin{equation}
F_{\mathrm{inc}}=\frac{\ln2}{\pi}\frac{P/f_{\mathrm{rep}}}{(x_{\mathrm{FWHM}}/2)(y_{\mathrm{FWHM}}/2)},
\end{equation}
where $P$, $f_{\mathrm{rep}}$, $x_{\mathrm{FWHM}}$, and $y_{\mathrm{FWHM}}$ are the average power, repetition rate, and horizontal and vertical spot diameters, respectively. One-minute power averages have standard errors below $2\%$.

\subsection{Beam-pointing and spot-size stability}

\begin{figure}[H]
    \centering
    \includegraphics[width=\linewidth]{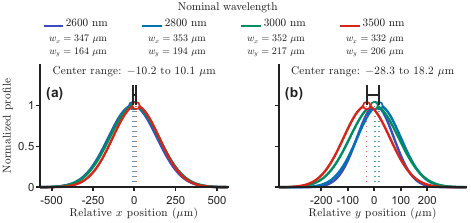}
    \caption{Representative knife-edge characterization of the DFG pump in the second measurement set at the nominal sample position. Normalized fitted Gaussian profiles are shown along the (a) horizontal and (b) vertical directions for nominal wavelength settings of $2600$, $2800$, $3000$, and $3500~\mathrm{nm}$. Positions are referenced to the fitted center at $3000~\mathrm{nm}$; dotted lines and open circles mark the fitted centers, and horizontal brackets indicate their full variation. The fitted FWHM values are given in the legends.}
    \label{pic:KE_size_and_positioning}
\end{figure}

We checked beam-pointing and spot-size stability during both measurement sets. Figure~\ref{pic:KE_size_and_positioning} shows a representative knife-edge series from the second set, recorded at the nominal sample position while sweeping the DFG wavelength without realigning the beam. Positions are referenced to the $3000$-nm setting.

Across the four wavelength settings, the fitted beam centers span $20.3~\mu\mathrm{m}$ horizontally and $46.5~\mu\mathrm{m}$ vertically, while the corresponding FWHM ranges are $332.0$--$353.1~\mu\mathrm{m}$ and $164.3$--$217.1~\mu\mathrm{m}$. Repeated scans at $3500~\mathrm{nm}$ after tuning away and back yield a horizontal-center standard deviation of $12.3~\mu\mathrm{m}$, providing an additional check of pointing reproducibility. The center excursions remain small relative to the pump widths, supporting stable pump--probe overlap during wavelength tuning.

\subsection{Pump spectral characterization}
\label{sec:pump_spectral_characterization}

\begin{figure}[H]
    \centering
    \includegraphics[width=0.98\linewidth]{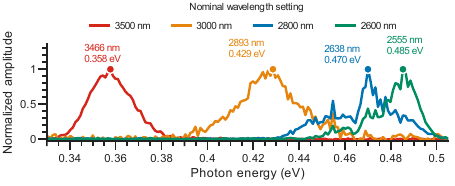}
    \caption{FTIR characterization of the DFG pump spectra used in the second measurement set, plotted against photon energy. The top legend gives the nominal wavelength settings; filled circles and adjacent labels give the measured maxima as both wavelength and photon energy. The spectra are normalized independently to their maxima. The measured peak energies $0.358$, $0.429$, $0.470$, and $0.485~\mathrm{eV}$ are used throughout the main text.}
    \label{pic:FTIR_pump}
\end{figure}

Figure~\ref{pic:FTIR_pump} illustrates the pump spectral characterization with representative Fourier-transform infrared (FTIR) spectra from the second measurement set. Here and below, we identify pump settings by their measured peak photon energies, consistent with the main text and facilitating comparison with the electronic band gap. The depletion near $0.45~\mathrm{eV}$ follows atmospheric water-vapor absorption~\cite{UCLAAtmosphericTransmission}, while the cutoff near $0.5~\mathrm{eV}$ reflects the DFG degeneracy limit. Assuming Gaussian spectral profiles, the root-mean-square (RMS) energy widths of the $0.358$- and $0.429$-eV pump spectra are approximately $0.0090$ and $0.0109~\mathrm{eV}$, respectively.

\section{Extraction of the coherent response}

\subsection{Timing convention and timing-jitter check}

Within each measurement set, the delay-stage calibration is kept fixed across pump photon energies. For traces at and below $0.429~\mathrm{eV}$, $t=0$ is defined by the first maximum of the transient response. The $0.470$- and $0.485$-eV responses have qualitatively different temporal profiles and are therefore not shifted independently; both retain the time zero established from the $0.429$-eV measurement. This convention preserves their delayed rises and relative phases.

As a timing sanity check, fluence-dependent traces at each photon energy up to $0.429~\mathrm{eV}$ are also shifted so that their first maxima coincide with that of the lowest-fluence trace. The inferred corrections remain below $100~\mathrm{fs}$ across the measured fluence range. An independent nearly constant-power energy sweep gives a comparable relative shift, approximately $100~\mathrm{fs}$ between the $0.358$- and $0.429$-eV settings. These checks bound externally introduced timing variations; they are not applied to erase the physically delayed rises at $0.470$ and $0.485~\mathrm{eV}$.

\subsection{Extraction of phonon oscillations}

\begin{figure}[H]
\centering
\includegraphics[width=\linewidth]{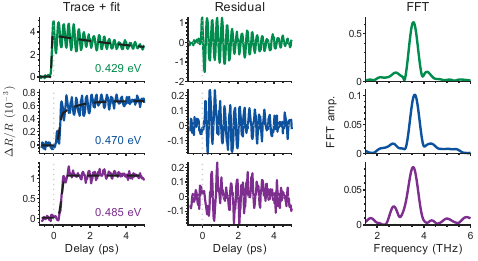}
\caption{Background subtraction for representative pump photon energies of $0.429$, $0.470$, and $0.485~\mathrm{eV}$ (top to bottom). The columns show the measured mean trace with its fitted background (black dashed), the residual, and its Hann-windowed Fourier amplitude spectrum over the nominal $0\leq t\leq5~\mathrm{ps}$ window.}
\label{fig:min_background}
\end{figure}

To extract the coherent-phonon signal reliably, we fit and subtract the non-oscillatory background in the time domain before Fourier analysis. A single functional form cannot adequately describe the background throughout the measured time window, so we construct $B(t)$ from piecewise fitting functions. Adjacent pieces are joined smoothly to avoid introducing artificial high-frequency components into the residual. Figure~\ref{fig:min_background} illustrates this procedure for three representative traces. All delays below are in picoseconds. The junctions use the quintic weight
\begin{equation}
s(u)=
\begin{cases}
0, & u\leq0,\\
10u^3-15u^4+6u^5, & 0<u<1,\\
1, & u\geq1,
\end{cases}
\label{eq:min_quintic_step}
\end{equation}
which satisfies
\begin{equation}
\begin{gathered}
s(0)=0,\qquad s(1)=1,\\
s'(0)=s'(1)=s''(0)=s''(1)=0.
\end{gathered}
\label{eq:min_quintic_constraints}
\end{equation}
For branches $L(t)$ and $R(t)$, define the connector
\begin{equation}
\mathcal C_{a,b}[L,R](t)=
\left[1-s\!\left(\frac{t-a}{b-a}\right)\right]L(t)
+s\!\left(\frac{t-a}{b-a}\right)R(t).
\label{eq:min_quintic_connector}
\end{equation}
Its endpoint constraints are explicitly
\begin{equation}
\begin{aligned}
\mathcal C_{a,b}^{(n)}(a)&=L^{(n)}(a),\\
\mathcal C_{a,b}^{(n)}(b)&=R^{(n)}(b),\qquad n=0,1,2.
\end{aligned}
\label{eq:min_connector_smoothness}
\end{equation}
Thus each join preserves the value, slope, and curvature. Between two constants the connector is a quintic polynomial; between a constant and an exponential it is a quintic-weighted blend.

\subsubsection{\normalfont\bfseries Self-similar traces: $\mathcal E\leq0.429~\mathrm{eV}$}

The traces in this range have self-similar prompt rises and slowly decaying backgrounds, exemplified by the $0.429$-eV trace in the top row of Fig.~\ref{fig:min_background}. We therefore use the same constant--quartic--exponential model for all photon energies and fluences in this range. Its fitted branches are
\begin{equation}
\begin{aligned}
P_4(t)&=C_{\mathrm{pre}}+\sum_{j=2}^{4}a_j(t+0.23)^j,\\
E(t)&=C_{\infty}+A\exp\!\left[-\frac{t-t_{\mathrm{ref}}}{\tau}\right].
\end{aligned}
\label{eq:min_bg_branches}
\end{equation}
The coefficients are fitted separately for each trace, with $C_{\mathrm{pre}}$ given by the mean over $t<-0.23$. The complete piecewise background, including the smooth joining intervals, is
\begin{equation}
B_{\leq0.429}(t)=
\begin{cases}
C_{\mathrm{pre}}, & t\leq t_1^-,\\
\mathcal C_{t_1^-,t_1^+}[C_{\mathrm{pre}},P_4](t),
 & t_1^-<t<t_1^+,\\
P_4(t), & t_1^+\leq t\leq t_2^-,\\
\mathcal C_{t_2^-,t_2^+}[P_4,E](t),
 & t_2^-<t<t_2^+,\\
E(t), & t\geq t_2^+.
\end{cases}
\label{eq:min_bg_low_energy}
\end{equation}
The interval $[t_1^-,t_1^+]$ joins the constant background to the quartic near the onset of the signal rise. The interval $[t_2^-,t_2^+]$ joins the quartic rise to the exponential decay near $t=0$, with $t_{\mathrm{ref}}=0$.

\subsubsection{\normalfont\bfseries Delayed rise: $\mathcal E=0.470~\mathrm{eV}$}

The $0.470$-eV trace in the middle row of Fig.~\ref{fig:min_background} shows a delayed rise toward a plateau. We join a constant early-time background smoothly to a saturating exponential:
\begin{equation}
B_{0.470}(t)=
\begin{cases}
C_{\mathrm{pre}}, & t\leq t_a,\\
\mathcal C_{t_a,t_b}[C_{\mathrm{pre}},E](t), & t_a<t<t_b,\\
E(t), & t\geq t_b.
\end{cases}
\label{eq:min_bg_post_0470}
\end{equation}
Here $C_{\mathrm{pre}}$ is the mean over $t\leq-0.10$, and the transition times are determined from each trace's smoothed rise. In the representative fit shown in Fig.~\ref{fig:min_background}, $t_a\simeq0.0611$, $t_b\simeq0.5444$, and $t_{\mathrm{ref}}=0.8$. The exponential remains the late-time branch throughout the analysis window.

\subsubsection{\normalfont\bfseries Persistent offset: $\mathcal E=0.485~\mathrm{eV}$}

The $0.485$-eV trace in the bottom row of Fig.~\ref{fig:min_background} develops a persistent offset with no resolved late-time decay. Both outer branches are therefore constant:
\begin{equation}
B_{0.485}(t)=
\begin{cases}
C_{\mathrm{pre}}, & t\leq0.16,\\
\mathcal C_{0.16,0.70}[C_{\mathrm{pre}},C_{\mathrm{late}}](t),
 & 0.16<t<0.70,\\
C_{\mathrm{late}}, & t\geq0.70.
\end{cases}
\label{eq:min_bg_0485}
\end{equation}
Here $C_{\mathrm{pre}}$ and $C_{\mathrm{late}}$ are the means over $t\leq0.16$ and $t\geq0.70$, respectively.

\subsubsection{\normalfont\bfseries Residual and Fourier spectrum}

The coherent residual is obtained from the raw transient reflectivity according to
\begin{equation}
\left(\frac{\Delta R}{R}\right)_{\mathrm{osc}}(t)
=\left(\frac{\Delta R}{R}\right)_{\mathrm{raw}}(t)-B(t).
\label{eq:min_residual}
\end{equation}
For every pump photon energy and fluence, the residual over the same time interval, $0\leq t\leq5~\mathrm{ps}$, is mean-subtracted, multiplied by a Hann window, and Fourier transformed. The same sampling interval, preprocessing, Fourier-transform normalization, and peak-extraction procedure are used throughout. The magnitude of the Fourier peak near the $A_1$ frequency of $3.6~\mathrm{THz}$ defines the $A_1$ amplitude. The second and third columns of Fig.~\ref{fig:min_background} show representative residual oscillations and their Fourier spectra, respectively.

\section{Optical parameters and phenomenological comparisons}

\subsection{Transmission-derived absorption and optical band gap}

To determine the optical absorption, we measured normal-incidence transmission through thin Te samples prepared with the same $c$-cut orientation and Czochralski growth method as the bulk crystal used for the pump--probe experiments. Their reduced thickness allowed measurable transmission near the band gap. The incident light propagated along the crystallographic $c$ axis. The measured transmission is
\begin{equation}
T(\mathcal{E})=\frac{I_{\mathrm{sam}}(\mathcal{E})}{I_{\mathrm{ref}}(\mathcal{E})},
\end{equation}
where $I_{\mathrm{sam}}$ and $I_{\mathrm{ref}}$ are the intensity spectra recorded with and without the sample, respectively. We subtract the instrumental leakage floor to obtain $T_{\mathrm{corr}}$ and retain only transmission points resolved above the noise floor.

For the measured thickness $D=164\pm4~\mu\mathrm{m}$, the absorption is extracted including multiple internal reflections. In the phase-averaged plane-parallel slab model, successive transmitted intensities form a geometric series. With $x=\exp[-\alpha_{\mathrm{eff}}(\mathcal E)D]$, this gives
\begin{equation}
T_{\mathrm{corr}}(\mathcal E)
=[1-R(\mathcal E)]^2x\sum_{m=0}^{\infty}[R(\mathcal E)^2x^2]^m
=\frac{[1-R(\mathcal E)]^2x}{1-R(\mathcal E)^2x^2}.
\label{eq:min_transmission_multiple_reflections}
\end{equation}
Inverting this relation yields
\begin{equation}
\alpha_{\mathrm{eff}}(\mathcal E)=\frac{1}{D}
\ln\!\left[
\frac{[1-R(\mathcal E)]^2+
\sqrt{[1-R(\mathcal E)]^4+4R(\mathcal E)^2T_{\mathrm{corr}}(\mathcal E)^2}}
{2T_{\mathrm{corr}}(\mathcal E)}
\right].
\label{eq:min_alpha}
\end{equation}
Here $R(\mathcal E)$ is the reflectivity interpolated from the optical constants of Ref.~\cite{palik1998handbook} and remains close to $0.44$ over the fitted range (see Table~\ref{tab:min_depth}). Internal reflections matter most where absorption is weak: compared with the single-pass extraction, $\alpha_{\mathrm{eff}}$ near $0.275~\mathrm{eV}$ increases from approximately $13.0$ to $19.5~\mathrm{cm}^{-1}$ (about $50\%$), whereas near $0.325~\mathrm{eV}$ it increases from approximately $82.3$ to $83.1~\mathrm{cm}^{-1}$ (less than $1\%$). The corresponding extinction coefficient is
\begin{equation}
k_{\mathrm{eff}}(\mathcal{E})=\frac{\alpha_{\mathrm{eff}}(\mathcal{E})\lambda}{4\pi}.
\label{eq:min_k_eff}
\end{equation}
The extracted subgap $k_{\mathrm{eff}}$ is approximately one order of magnitude larger than the literature value~\cite{palik1998handbook}, possibly reflecting sample-dependent doping, defects, or residual scattering losses. The finite absorption at the lowest pump photon energy, $0.275~\mathrm{eV}$, therefore permits weak linear absorption below the gap.

\begin{figure}[tbp]
\centering
\includegraphics[width=0.80\linewidth]{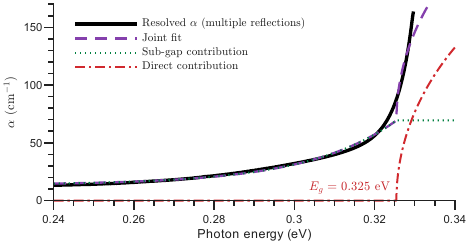}
\caption{Determination of the optical band gap from the FTIR transmission spectrum. The black curve is the effective absorption coefficient extracted with multiple internal reflections included. The dashed magenta curve is the joint fit of Eq.~\eqref{eq:min_bandgap_fit}, decomposed into the dotted green Urbach-like subgap contribution and the dash-dotted red direct-allowed contribution. The fitted gap is $E_g=0.325~\mathrm{eV}$.}
\label{fig:min_bandgap}
\end{figure}

Figure~\ref{fig:min_bandgap} shows a gradual subgap increase followed by a much sharper absorption onset. To separate these contributions phenomenologically, we fit the corrected absorption with
\begin{equation}
\alpha_{\mathrm{fit}}(\mathcal{E})=
\alpha_0+
\alpha_{\mathrm{U},g}
\exp\!\left[
\frac{\min(\mathcal{E},E_g)-E_g}{E_{\mathrm U}}
\right]
+
\frac{A}{\mathcal{E}}
\sqrt{\max(\mathcal{E}-E_g,0)}.
\label{eq:min_bandgap_fit}
\end{equation}
The terms describe an energy-independent background, an Urbach-like tail, and a direct-allowed onset for parabolic bands, respectively~\cite{urbach1953long,pankove1971optical,ferlauto2002analytical}. Here $\alpha_{\mathrm{U},g}$ is the tail contribution at the gap, $E_{\mathrm U}$ is its characteristic energy, and $A$ is the direct-transition amplitude. Holding the tail constant above $E_g$ prevents its extrapolated exponential growth from obscuring the interband contribution.

The fit yields $E_g\simeq0.325~\mathrm{eV}$ and it agrees with the room-temperature perpendicular-polarization edge of $0.324~\mathrm{eV}$ and the established bulk gap near $0.33~\mathrm{eV}$~\cite{loferski1954infrared,tutihasi1969optical,anzin1977measurement}. Transmission reaches the noise floor shortly above the edge, limiting the available above-gap range and preventing a stringent independent test of the square-root line shape. The extracted gap should therefore be interpreted as a phenomenological onset estimate rather than a precision determination of the band dispersion.

\subsection{Combined optical constants and surface carrier density}

\begin{figure}[tbp]
    \centering
    \includegraphics[width=\linewidth]{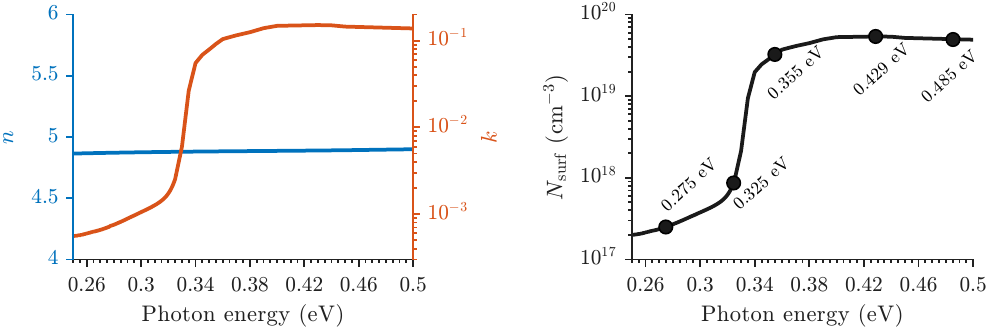}
    \caption{Optical parameters used in the analysis. (a) Complex in-plane refractive index $\widetilde n=n+\mathrm{i}k$ near the fundamental band gap. The refractive index $n$ and the above-gap values of $k$ are taken from Ref.~\cite{palik1998handbook}, while the subgap values of $k$ are extracted from the measured transmission spectrum including multiple internal reflections. (b) Estimated surface photoexcited-carrier density $N_{\mathrm{surf}}$ for $F_{\mathrm{inc}}=1~\mathrm{mJ\,cm^{-2}}$. Markers indicate selected experimental pump photon energies.}
    \label{fig:min_optical_density}
\end{figure}

To incorporate the sample-specific subgap absorption into the analysis, we construct a combined optical dataset: the refractive index $n$ and the above-gap extinction coefficient $k$ are taken from Ref.~\cite{palik1998handbook}, whereas for $\mathcal{E}\leq0.325~\mathrm{eV}$ we use the transmission-derived $k_{\mathrm{eff}}=\alpha_{\mathrm{eff}}\lambda/(4\pi)$ of Eq.~\eqref{eq:min_k_eff}. All optical quantities below are evaluated from this combined dataset. Figure~\ref{fig:min_optical_density}(a) shows the resulting $\widetilde n=n+\mathrm{i}k$: the real part of the dielectric function remains close to $24$ across the band gap, while the imaginary part grows by more than two orders of magnitude but remains much smaller.

The relevant length scale for axial non-uniformity is the optical penetration depth, defined as the $1/e$ decay length of the intensity in the Beer--Lambert limit,
\begin{equation}
d(\mathcal{E})=\frac{1}{\alpha(\mathcal{E})}=\frac{\lambda}{4\pi k(\mathcal{E})}.
\label{eq:min_pen_depth}
\end{equation}
Table~\ref{tab:min_depth} lists $d$ at representative pump and probe wavelengths. Even at the shortest pump wavelength ($2555~\mathrm{nm}$, $0.485~\mathrm{eV}$), the pump penetrates approximately $24$ times farther than the $800$-nm probe, so $d_{\mathrm{probe}}\ll d_{\mathrm{pump}}$ throughout the investigated range and the excitation is approximately uniform over the probed depth. 

\begin{table}[tbp]
\caption{Representative pump and probe wavelengths, photon energies, optical penetration depths defined by Eq.~\eqref{eq:min_pen_depth}, and reflectivities, all calculated from the combined optical dataset. Penetration depths marked by $\dagger$ are extracted from the measured transmission including multiple internal reflections; their reflectivities are evaluated using the corresponding corrected $k$ and literature $n$ at the listed wavelengths; the remaining values are calculated from literature optical constants~\cite{palik1998handbook}.}
\label{tab:min_depth}
\centering
\begin{ruledtabular}
\begin{tabular}{cccc}
Wavelength (nm) & Photon energy (eV) & $d$ ($\mu\mathrm{m}$) & $R$ \\
\hline
4511 & 0.275 & $513^{\dagger}$ & 0.435 \\
3820 & 0.325 & $126^{\dagger}$ & 0.436 \\
3495 & 0.355 & 3.00 & 0.440 \\
2893 & 0.429 & 1.52 & 0.440 \\
2555 & 0.485 & 1.46 & 0.440 \\
800 (probe) & 1.55 & 0.061 & 0.512 \\
\end{tabular}
\end{ruledtabular}
\end{table}

Assuming linear absorption and one electron--hole pair generated per absorbed photon, the surface carrier density is
\begin{equation}
N_{\mathrm{surf}}=\frac{F_{\mathrm{inc}}[1-R]\alpha(\mathcal{E})}{\mathcal{E}},
\label{eq:min_nsurf}
\end{equation}
where $F_{\mathrm{inc}}$ is the incident fluence. Fig.~\ref{fig:min_optical_density}(b) shows Eq.~\eqref{eq:min_nsurf} for $F_{\mathrm{inc}}=1~\mathrm{mJ\,cm^{-2}}$; the main text compares its rescaled spectral dependence with the $A_1$ responsivity.

\section{Two-Raman-tensor force analysis}

We compare the measured photon-energy dependence with the two-Raman-tensor description of coherent-phonon generation introduced by Stevens \textit{et al.}~\cite{stevens2002coherent}. For a fixed pump-polarization channel and photon energy $\mathcal{E}=\hbar\omega$, the two-Raman-tensor force is
\begin{equation}
F_{\mathrm{2R}}(\Omega;\mathcal{E})\propto
\left[
\frac{d\epsilon_1}{d\mathcal{E}}+
\frac{2\mathrm{i}\epsilon_2}{\hbar\Omega}
\right]\widetilde I(\Omega),
\label{eq:min_stevens_force}
\end{equation}
where $\epsilon=\epsilon_1+\mathrm{i}\epsilon_2$ and
\begin{equation}
\widetilde I(\Omega)=\int_{-\infty}^{\infty}
e^{\mathrm{i}\Omega t}|E(t)|^2\,dt
\label{eq:min_intensity_fourier}
\end{equation}
is the Fourier component of the pump-intensity envelope at the $A_1$ angular frequency $\Omega$. The factor arising from $d/d\omega=\hbar\,d/d\mathcal{E}$ is absorbed into the overall proportionality constant. The first term in Eq.~\eqref{eq:min_stevens_force} is dispersive, whereas $2\epsilon_2/(\hbar\Omega)$ is the direct absorptive contribution associated with real electronic excitation.

Equation~\eqref{eq:min_stevens_force}, which is the form given by Eqs.~(9) and (10) of Ref.~\cite{stevens2002coherent}, retains only the real part of the Raman derivative; the imaginary part, $\mathrm{i}\,d\epsilon_2/d\mathcal{E}$, is omitted there. Because $\epsilon_2$ varies rapidly across the Te band gap, this term is not negligible for the present measurements, and we therefore evaluate the extended form that retains the full complex derivative,
\begin{equation}
F_{\mathrm{2R}}(\Omega;\mathcal{E})\propto
\left[
\frac{d\epsilon_1}{d\mathcal{E}}+
\mathrm{i}\frac{d\epsilon_2}{d\mathcal{E}}+
\frac{2\mathrm{i}\epsilon_2}{\hbar\Omega}
\right]\widetilde I(\Omega).
\label{eq:min_two_raman_local}
\end{equation}
The dielectric function is calculated from the combined optical constants described above; the measured subgap and literature above-gap datasets are smoothly interpolated through their joining region before differentiation to suppress numerical artifacts.

The finite pump bandwidth is included by representing the spectral intensity as a normalized Gaussian $w_\tau(\mathcal{E};\mathcal{E}_0)$ centered at $\mathcal{E}_0$,
\begin{equation}
\left\langle g\right\rangle_{\mathcal{E}_0,\tau}
=\int d\mathcal{E}\,w_\tau(\mathcal{E};\mathcal{E}_0)g(\mathcal{E}),
\qquad
\int d\mathcal{E}\,w_\tau(\mathcal{E};\mathcal{E}_0)=1,
\label{eq:min_two_raman_average}
\end{equation}
where $\tau$ is the transform-limited intensity FWHM. The bandwidth-averaged force used for comparison is therefore
\begin{equation}
\begin{aligned}
F_{\mathrm{2R}}(\Omega;\mathcal{E}_0,\tau)\propto \widetilde{I}_{\mathrm{abs},\tau}(\Omega;\mathcal{E}_0)
\Bigg[
\left\langle\frac{d\epsilon_1}{d\mathcal{E}}\right\rangle_{\mathcal{E}_0,\tau}+\mathrm{i}\Bigg(
\left\langle\frac{d\epsilon_2}{d\mathcal{E}}\right\rangle_{\mathcal{E}_0,\tau}
+\frac{2}{\hbar\Omega}\left\langle\epsilon_2\right\rangle_{\mathcal{E}_0,\tau}
\Bigg)\Bigg].
\end{aligned}
\label{eq:min_two_raman}
\end{equation}
Both $w_\tau$ and $\widetilde I_{\mathrm{abs},\tau}$ are evaluated consistently for Gaussian pulses of the stated duration and a common incident fluence. Because $R$ varies only weakly across the experimental range (Table~\ref{tab:min_depth}) and every measurement uses the same $800$-nm probe, both the absorbed-envelope factor and the optical detection tensor contribute only a common scale, independent of central pump energy. 

\begin{figure}[tbp]
\centering
\includegraphics[width=\linewidth]{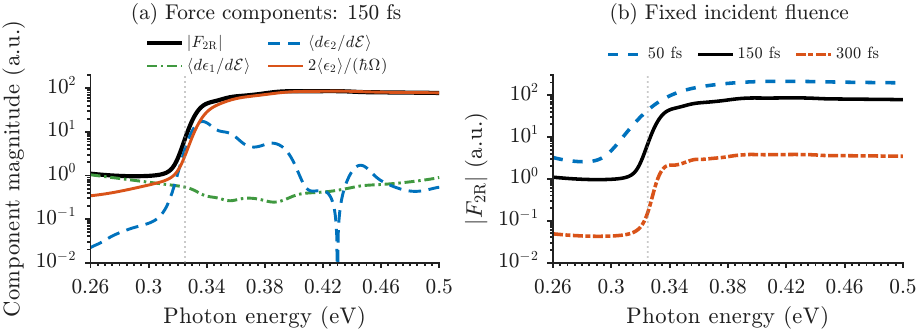}
\caption{Two-Raman-tensor force near the Te band gap. (a) Magnitudes of the bandwidth-averaged dispersive, derivative-absorptive, and direct absorptive contributions in Eq.~\eqref{eq:min_two_raman}, and the total force magnitude, for a $150$-fs transform-limited pulse. (b) Total force magnitude for $50$, $150$, and $300$-fs transform-limited pulses at equal incident fluence. All curves share a common scale, defined by the $150$-fs total force at $0.275~\mathrm{eV}$. Vertical dotted lines mark $E_g\simeq0.325~\mathrm{eV}$.}
\label{fig:min_two_raman}
\end{figure}

Fig.~\ref{fig:min_two_raman}(a) shows that the subgap response is governed mainly by $d\epsilon_1/d\mathcal{E}$. Once interband absorption becomes appreciable, $2\epsilon_2/(\hbar\Omega)$ dominates and raises the predicted force by nearly two orders of magnitude. Figure~\ref{fig:min_two_raman}(b) shows that pulse duration changes the overall force magnitude and the bandwidth-induced smoothing of the gap onset, while the large subgap-to-above-gap contrast persists. Thus the two-Raman-tensor model predicts the opposite of the measured reduction in $A_1$ responsivity and cannot account for it through finite bandwidth, reflectivity, or probe detection.

\section{Quasi-equilibrium constrained-DFT calculation}

\subsection{Computational framework}

First-principles calculations are performed with the plane-wave pseudopotential implementation of density functional theory in \textsc{Quantum ESPRESSO}~\cite{giannozzi2009quantum,giannozzi2017advanced,giannozzi2020quantum}. Core--valence interactions are described by the scalar-relativistic projector-augmented-wave pseudopotential \path{Te.pbe-n-kjpaw_psl.1.0.0.UPF} from PSLibrary~\cite{dalcorso2014pseudopotentials}, together with the Perdew--Burke--Ernzerhof generalized-gradient approximation~\cite{perdew1996generalized}. Wave-function and charge-density cutoffs of $60$ and $240~\mathrm{Ry}$ are used, respectively, and Brillouin-zone integrations employ a $24\times24\times20$ $k$-point mesh.

Trigonal Te in space group $P3_1 21$ is first relaxed with respect to both the lattice parameters and internal coordinates. Using a self-consistent relaxed structure is important because the experimental coordinates need not coincide with the minimum predicted by the chosen exchange--correlation functional~\cite{tangney1999calculations,tangney2002density}. The resulting reference structure has $a=4.503~\text{\AA}$, $c=5.959~\text{\AA}$, and dimensionless internal Wyckoff parameter $x_{\mathrm{Te},0}=0.2708$.

We denote the unweighted Cartesian amplitude of the three-atom $A_1$ displacement by $u_{A_1}$, measured in units of length. With a unit-normalized Cartesian eigendisplacement vector, the displacement relative to the reference structure is decomposed as
\begin{equation}
\Delta\mathbf{R}=\mathbf{R}-\mathbf{R}_0
=u_{A_1}\mathbf{e}_{A_1}+\Delta\mathbf{R}_{\perp},
\label{eq:min_A1_decomposition}
\end{equation}
where $\mathbf{R}$ and $\mathbf{R}_0$ concatenate the Cartesian coordinates of the three atoms in the primitive cell. The normalized eigenvector satisfies $\mathbf{e}_{A_1}\cdot\mathbf{e}_{A_1}=1$, so that
\begin{equation}
u_{A_1}=\Delta\mathbf{R}\cdot\mathbf{e}_{A_1},
\qquad
\Delta\mathbf{R}_{\perp}=\Delta\mathbf{R}-u_{A_1}\mathbf{e}_{A_1}.
\label{eq:min_A1_projection}
\end{equation}
The coordinate $u_{A_1}$ uses the Euclidean projection of the Cartesian displacements, with no mass factor. For a pure $A_1$ displacement, each Te atom moves a distance $|u_{A_1}|/\sqrt{3}$. In the Cartesian convention used here, the three atomic components are
\begin{equation}
\begin{aligned}
\mathbf{e}_1&=\left(\frac{1}{2\sqrt{3}},-\frac{1}{2},0\right),&
\mathbf{e}_2&=\left(\frac{1}{2\sqrt{3}},+\frac{1}{2},0\right),\\
\mathbf{e}_3&=\left(-\frac{1}{\sqrt{3}},0,0\right).
\end{aligned}
\label{eq:min_A1_eigenvector}
\end{equation}
The eigenvector phase is chosen such that positive $u_{A_1}$ denotes expansion of the Te helix and negative $u_{A_1}$ denotes contraction.

\subsection{Quasi-equilibrium photoexcited structural relaxation}

The quasi-equilibrium displacement is estimated using the two-chemical-potential method implemented through \texttt{twochem = .true.}~\cite{marini2021lattice}. A prescribed number of electrons, controlled by \texttt{nelec\_cond}, is transferred from the valence to the conduction manifold while maintaining overall charge neutrality. Electron and hole occupations are represented by separate Fermi--Dirac distributions at a common effective carrier temperature; $k_{\mathrm B}T=0.05~\mathrm{eV}$ is used. This construction describes a partially thermalized, charge-neutral electron--hole plasma and is distinct from the transition-selective population immediately after optical excitation.

Excitation fractions from $0.001\%$ to $1\%$ of the valence-electron population are considered. Using the Te valence-electron density $1.77\times10^{23}~\mathrm{cm}^{-3}$, this corresponds to carrier densities between $1.77\times10^{18}$ and $1.77\times10^{21}~\mathrm{cm}^{-3}$. At each density, the lattice parameters are held at their ground-state values and the internal coordinates are relaxed with the BFGS algorithm until the force and total-energy thresholds reach $10^{-5}~\mathrm{Ry/Bohr}$ and $10^{-5}~\mathrm{Ry}$, respectively. These calculations therefore determine the minimum of the quasi-equilibrium lattice potential for a prescribed thermalized carrier population.

Each relaxed displacement is projected using Eq.~\eqref{eq:min_A1_projection}. The orthogonal residual is quantified by
\begin{equation}
\mathrm{RMS}_{\perp}=\sqrt{\frac{1}{3N}\sum_{i,\alpha}
\left(\Delta R_{\perp,i\alpha}\right)^2},
\qquad N=3,
\label{eq:min_A1_residual}
\end{equation}
where $i$ labels atoms and $\alpha\in\{x,y,z\}$. We obtain $\mathrm{RMS}_{\perp}<10^{-9}~\text{\AA}$ throughout, confirming that the relaxed structures remain in the $A_1$ symmetry subspace to numerical precision.

\begin{figure}[tbp]
\centering
\includegraphics[width=0.88\linewidth]{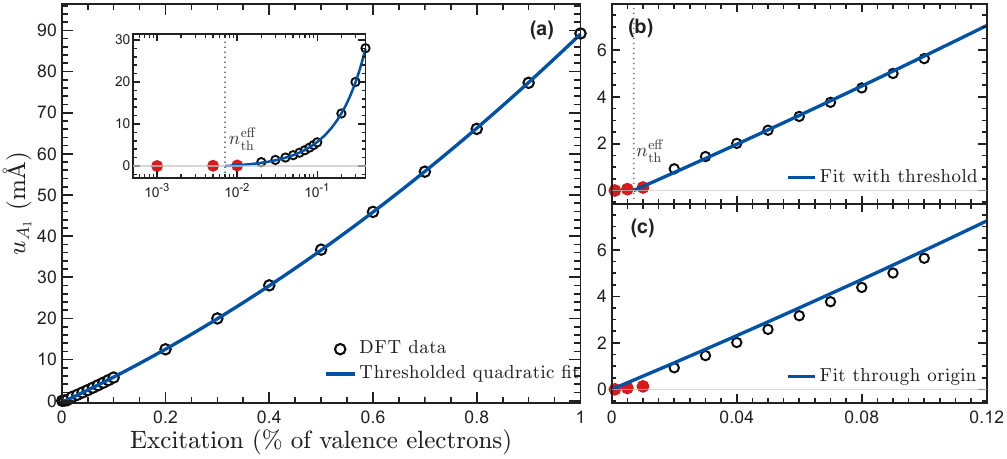}
\caption{Calculated quasi-equilibrium displacement $u_{A_1}$ versus excitation fraction. Open circles are constrained-DFT results; curves show the empirical thresholded quadratic fit and, for comparison, a quadratic fit constrained to pass through the origin. The low-density panels enlarge the experimentally relevant regime.}
\label{fig:min_Q}
\end{figure}

Over the calculated range, the density dependence is represented by the empirical thresholded quadratic form
\begin{equation}
u_{A_1}(n)=31\widetilde n^{\:2}+59\widetilde n,
\qquad
\widetilde n=\max(n-n_{\mathrm{th}}^{\mathrm{eff}},0),
\qquad
n_{\mathrm{th}}^{\mathrm{eff}}\simeq0.007\%,
\label{eq:min_thresholded_Q}
\end{equation}
where $u_{A_1}$ is in $\mathrm{m}\text{\AA}$ and $n$ is the numerical excitation fraction expressed in percent. A quadratic constrained to the origin gives $u_{A_1}(n)=33n^2+56n$ but systematically overestimates the resolved low-density displacements. Since $n_{\mathrm{th}}^{\mathrm{eff}}$ lies where the relaxed displacement approaches the numerical resolution, it is an effective numerical onset rather than evidence for a sharp physical threshold. The fits are used only to interpolate the quasi-equilibrium displacement into the experimentally relevant low-density regime.

\subsection{Benchmark against previous first-principles results}

\begin{figure}[tbp]
    \centering
    \includegraphics[width=0.80\linewidth]{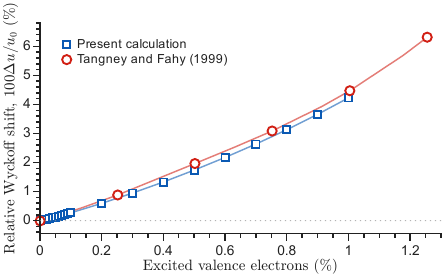}
    \caption{Benchmark comparison of the photoinduced internal-coordinate change obtained here with the first-principles results of Tangney and Fahy~\cite{tangney1999calculations}. Their values are digitized from the published figure and referenced to the reported equilibrium coordinate $x_{\mathrm{eq}}=0.2686$. The present Cartesian amplitude is converted using $\Delta x_{\mathrm{Te}}=u_{A_1}/(\sqrt{3}a)$, with $a=4.503~\text{\AA}$, and referenced to $x_{\mathrm{Te},0}=0.2708$. The vertical axis shows the percentage change relative to the independently calculated equilibrium coordinate of each dataset. Lines are guides to the eye.}
    \label{fig:min_Wyckoff_comparison}
\end{figure}

To validate the calculated structural response, we benchmark it against the independent first-principles results of Tangney and Fahy~\cite{tangney1999calculations}, which address the same photoinduced expansion of the Te helix. The two studies express this response in different coordinates: Ref.~\cite{tangney1999calculations} reports the internal Wyckoff coordinate rather than the unweighted Cartesian amplitude $u_{A_1}$ defined above. Since $\Delta x_{\mathrm{Te}}=x_{\mathrm{Te}}-x_{\mathrm{Te},0}$ moves each Te atom by $a|\Delta x_{\mathrm{Te}}|$ in the basal plane,
\begin{equation}
u_{A_1}=\sqrt{3}\,a\,\Delta x_{\mathrm{Te}},
\qquad
\Delta x_{\mathrm{Te}}=\frac{u_{A_1}}{\sqrt{3}\,a}.
\label{eq:min_Q_to_Wyckoff}
\end{equation}
Because the two calculations relax to slightly different equilibrium coordinates ($x_{\mathrm{Te},0}=0.2708$ here and $x_{\mathrm{eq}}=0.2686$ in Ref.~\cite{tangney1999calculations}), we compare relative changes,
\begin{equation}
\left(\frac{\Delta x_{\mathrm{Te}}}{x_{\mathrm{Te},0}}\right)_{\mathrm{present}}
=\frac{u_{A_1}}{\sqrt{3}\,a\,x_{\mathrm{Te},0}},
\qquad
\left(\frac{\Delta x}{x_{\mathrm{eq}}}\right)_{\mathrm{TF}}
=\frac{x-x_{\mathrm{eq}}}{x_{\mathrm{eq}}}.
\label{eq:min_relative_Wyckoff}
\end{equation}
The literature values are digitized without further shifting or rescaling. Referencing each displacement to its own ground state removes this method-dependent offset while preserving the magnitude and density dependence of the response. Figure~\ref{fig:min_Wyckoff_comparison} shows that the two datasets agree closely in both, despite the differences in computational details and structural conventions. This agreement confirms that the calculated expansion of the Te helix is consistent with the established first-principles result, and it is this benchmarked response that underlies the experimental comparison in the main text.

\section{Transition-resolved prethermal force}

\subsection{Force generated by a nonequilibrium occupation}

For a prescribed set of instantaneous electronic occupations $\{f_{n\mathbf{k}}(t)\}$, the generalized force conjugate to the normal coordinate $Q_{A_1}$ is
\begin{equation}
\mathcal{F}_{A_1}\!\left[\{f_{n\mathbf{k}}\}\right]=-
\left.\frac{\partial E_{\mathrm{BO}}\!\left[\{f_{n\mathbf{k}}\},Q_{A_1}\right]}{\partial Q_{A_1}}\right|_{Q_{A_1}=0},
\label{eq:min_force_general}
\end{equation}
where $E_{\mathrm{BO}}$ is the Born--Oppenheimer energy per cell. The ions remain at the ground-state structure, and the occupations are held fixed under the derivative, $\left.\partial f_{n\mathbf{k}}/\partial Q_{A_1}\right|_{Q_{A_1}=0}=0$. The equilibrium force vanishes at the relaxed structure. To first order in $\delta f_{n\mathbf{k}}=f_{n\mathbf{k}}-f^{(0)}_{n\mathbf{k}}$ within the weak-excitation, frozen-band limit,
\begin{equation}
\mathcal{F}_{A_1}(t)\simeq-\sum_{n,\mathbf{k}}\delta f_{n\mathbf{k}}(t)g_n(\mathbf{k}),
\label{eq:min_force_linearized}
\end{equation}
where the Brillouin-zone weights are included in the $\mathbf{k}$ sum and $g_n(\mathbf{k})=\left.\partial E_{n\mathbf{k}}/\partial Q_{A_1}\right|_{Q_{A_1}=0}$ is the band- and momentum-resolved electron--phonon coupling to $Q_{A_1}$.

\subsection{Rewriting the force as a sum over optical transitions}

For a vertical transition $v\rightarrow c$ at fixed $\mathbf{k}$, let $\delta f_{cv}$ be the transferred population. Pair conservation gives
\begin{equation}
\delta f_{c\mathbf{k}}=\sum_v\delta f_{cv}(\mathbf{k}),
\qquad
\delta f_{v\mathbf{k}}=-\sum_c\delta f_{cv}(\mathbf{k}).
\label{eq:min_population_conservation}
\end{equation}
Substitution into Eq.~\eqref{eq:min_force_linearized} pairs the electron and hole contributions,
\begin{equation}
\mathcal{F}_{A_1}(t)=-\sum_{\mathbf{k},v,c}\delta f_{cv}(\mathbf{k},t)\Delta g_{cv}(\mathbf{k}),
\label{eq:min_force_transition}
\end{equation}
with
\begin{equation}
\begin{split}
\Delta g_{cv}(\mathbf{k})=g_c(\mathbf{k})-g_v(\mathbf{k})=\left.\frac{\partial\Delta E_{cv}(\mathbf{k})}{\partial Q_{A_1}}\right|_{Q_{A_1}=0},
\end{split}
\qquad
\Delta E_{cv}(\mathbf{k})=E_{c\mathbf{k}}-E_{v\mathbf{k}}.
\label{eq:min_transition_coupling}
\end{equation}
Each pair is therefore governed by the deformation potential of its own transition energy, whose sign sets its force direction.

\subsection{Optically generated transition population}

In the weak-excitation regime, before significant state filling or carrier redistribution occurs, first-order time-dependent perturbation theory gives the post-pulse population transfer as
\begin{equation}
\delta f_{cv}(\mathbf{k};\mathcal{E})
\propto F_{\mathrm{inc}}
\left|M_{cv}(\mathbf{k})\right|^2
G\!\left[\mathcal{E}-\Delta E_{cv}(\mathbf{k}),\sigma\right],
\label{eq:min_population_ansatz}
\end{equation}
where $M_{cv}(\mathbf{k})=\mathbf{e}\cdot\mathbf{r}_{cv}(\mathbf{k})$ is the matrix element for pump polarization $\mathbf{e}$ and interband position element $\mathbf{r}_{cv}$. The normalized Gaussian
\begin{equation}
G(x,\sigma)=\frac{1}{\sigma\sqrt{2\pi}}
\exp\!\left(-\frac{x^2}{2\sigma^2}\right)
\label{eq:min_gaussian}
\end{equation}
replaces the energy-conservation delta function. Here $\sigma$ is an effective broadening parameter that combines the pump bandwidth and electronic broadening.

Combining Eqs.~\eqref{eq:min_force_transition} and \eqref{eq:min_population_ansatz} yields the transition-resolved prethermal-force spectrum used in the main text,
\begin{equation}
\begin{aligned}
\mathcal{F}_{A_1}(\mathcal{E})
&\propto-F_{\mathrm{inc}}\sum_{\mathbf{k},v,c}
\Delta g_{cv}(\mathbf{k})\\
&\quad\times \left|M_{cv}(\mathbf{k})\right|^2 G\!\left[\mathcal{E}-\Delta E_{cv}(\mathbf{k}),\sigma\right].
\end{aligned}
\label{eq:min_force}
\end{equation}

Removing the signed deformation-potential factor defines the dipole-weighted joint density of states (JDOS),
\begin{equation}
\mathcal{J}(\mathcal{E})\propto
\sum_{\mathbf{k},v,c}\left|M_{cv}(\mathbf{k})\right|^2
G\!\left[\mathcal{E}-\Delta E_{cv}(\mathbf{k}),\sigma\right].
\label{eq:min_dipole_weighted_jdos}
\end{equation}

\subsection{Numerical evaluation, symmetry, and sign convention}

The deformation potentials are evaluated by symmetric frozen-phonon differences at $Q_{A_1}=\pm\delta Q$, with $\delta Q=0.004~\text{\AA}$,
\begin{equation}
g_n(\mathbf{k})\simeq
\frac{E_{n\mathbf{k}}(+\delta Q)-E_{n\mathbf{k}}(-\delta Q)}{2\delta Q}.
\label{eq:min_finite_difference}
\end{equation}
This requires three independent electronic-structure calculations: one for the equilibrium structure and one for each of the two displaced structures. All three use identical pseudopotentials, plane-wave cutoffs, band windows, and Wannier projections. Each is Wannier-interpolated separately~\cite{pizzi2020wannier90}, so that the force and JDOS spectra can be evaluated on a dense uniform $\mathbf k$ mesh. Because Eq.~\eqref{eq:min_finite_difference} differences the eigenvalues of the same band in the two displaced structures, the band indices of the three interpolations must be brought into one-to-one correspondence, otherwise the difference would mix different physical bands. The bands are therefore matched by eigenvalue continuity, so that $n$ labels the same physical band in all three structures. A rigid $0.115$-eV scissor shift applied to the conduction bands aligns the $0.21$-eV PBE gap with the experimental $0.325~\mathrm{eV}$. Being independent of $Q_{A_1}$, the shift enters $\Delta E_{cv}$ but cancels from $g_n$ and $\Delta g_{cv}$. The sums run over the three highest valence bands and the three lowest conduction bands, which together span transition energies up to $\hbar\omega\simeq5~\mathrm{eV}$, far above the $0.5~\mathrm{eV}$ upper limit of the range investigated here.

For clarity, the polarization dependence can be written as an in-plane symmetric tensor,
\begin{equation}
\mathcal{F}_{\alpha\beta}(\mathcal{E})\propto
-F_{\mathrm{inc}}\sum_{\mathbf{k},v,c}\Delta g_{cv}(\mathbf{k})
\operatorname{Re}\!\left[r_{cv}^{\alpha}(\mathbf{k})r_{cv}^{\beta *}(\mathbf{k})\right]
G\!\left[\mathcal{E}-\Delta E_{cv}(\mathbf{k}),\sigma\right],
\label{eq:min_force_tensor}
\end{equation}
where $\alpha,\beta\in\{x,y\}$.  Thus, for an in-plane linear polarization $\mathbf e=(\cos\theta,\sin\theta)$, the raw result is
\begin{equation}
\mathcal F_{A_1}^{\mathrm{raw}}(\mathcal E,\theta)=
\mathcal F_{xx}\cos^2\theta+\mathcal F_{yy}\sin^2\theta
+2\mathcal F_{xy}\sin\theta\cos\theta.
\label{eq:min_force_angular}
\end{equation}
Because an $A_1$ displacement preserves the $D_3$ point group, the exact in-plane response must be rotationally invariant: $\mathcal F_{xx}=\mathcal F_{yy}$ and $\mathcal F_{xy}=0$. We therefore use the angular variation of Eq.~\eqref{eq:min_force_angular} as a numerical symmetry check, rather than assuming invariance at the outset.

\begin{figure}[tbp]
\centering
\includegraphics[width=\linewidth]{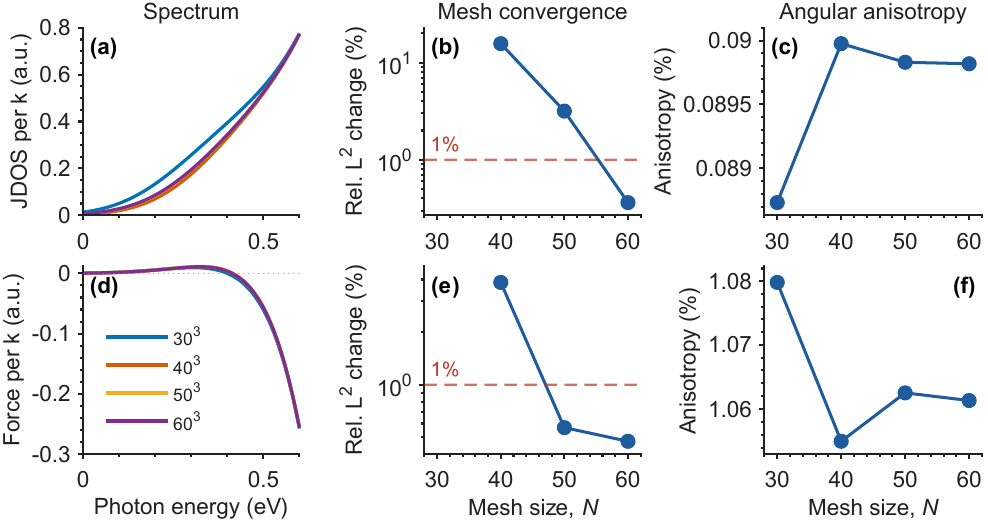}
\caption{Convergence of the dipole-weighted JDOS (top row) and prethermal force (bottom row) with interpolation-mesh density for $\sigma=0.15~\mathrm{eV}$, together with the residual in-plane anisotropy before $D_3$ projection. The dashed horizontal lines mark the $1\%$ spectral-convergence criterion.}
\label{fig:min_convergence}
\end{figure}

Figure~\ref{fig:min_convergence} shows the mesh convergence of the dipole-weighted JDOS and force spectra, together with the check of rotational invariance described above. Because the force crosses zero, convergence for each $\sigma$ is quantified using the relative $L^2$ difference between successive spectral traces, avoiding unstable pointwise ratios. At $\sigma=0.15~\mathrm{eV}$, for example, the force changes by less than $1\%$ between the $40^3$ and $50^3$ meshes and again between the $50^3$ and $60^3$ meshes, whereas the JDOS reaches the same criterion only on the $60^3$ mesh.

The residual peak-to-peak angular anisotropy is approximately $1.06\%$ for the raw force and $0.09\%$ for the JDOS, and it does not decrease systematically with mesh density. We therefore attribute it to a small numerical residual of the interpolation and of the independently generated Wannier representations, rather than to a physical anisotropy. Because this residual is numerical, the reported spectra are projected onto the $D_3$-symmetric component,
\begin{equation}
\mathcal{F}_{\parallel}^{\mathrm{sym}}(\mathcal{E})=
\frac{\mathcal{F}_{xx}(\mathcal{E})+\mathcal{F}_{yy}(\mathcal{E})}{2},
\qquad \mathcal{F}_{xy}^{\mathrm{sym}}=0,
\label{eq:min_D3_projection}
\end{equation}
which removes any dependence on the in-plane polarization angle. The dipole-weighted JDOS is projected in the same manner. These averaged spectra are the prethermal force and JDOS shown in the main text. Eqs.~\eqref{eq:min_force_general}--\eqref{eq:min_D3_projection} use $\mathcal F_{A_1}=-\partial E_{\mathrm{BO}}/\partial Q_{A_1}$ with positive $Q_{A_1}$ denoting expansion. Fig.~4 of the main text and Fig.~\ref{fig:min_convergence} here reverse the plotted sign so positive denotes contraction.

\subsection{Fitting procedure}

The experiment fixes the magnitude of the coherent oscillation but not its absolute sign, so the scale factor is left free in sign. For each $\sigma$, it is fitted to the entire calculated spectrum by least squares over $0.32$--$0.475~\mathrm{eV}$ on the mesh converged for that broadening while the dipole-weighted JDOS is evaluated with the same $\sigma$. Points below the gap are excluded, since the model describes only real interband transitions, and the $0.485$-eV point is excluded because a single measurement cannot identify the linear fluence regime; it is retained as an independent comparison near the calculated reversal. The RMS residual rises on both sides of a minimum at $\sigma=0.125~\mathrm{eV}$, and since the scale is the only fitted quantity, the high-energy suppression, the zero crossing, and the sign reversal are predictions rather than fit constraints. 

\section{Data and code availability}

The data and analysis scripts supporting this study are available from the corresponding authors upon reasonable request.

\clearpage
\bibliography{references_PRL_cleaned}
\bibliographystyle{apsrev4-2}